\documentclass[doublecol]{epl2} 

\usepackage{amsmath}
\usepackage{cleveref}
\usepackage{subcaption}

\usepackage{enumerate}
\usepackage{microtype}

\title{The Nonequilibrium Crystallization Force}
\shorttitle{Nonequilibrium Crystallization Force} 

\author{Luca Gagliardi \and Olivier Pierre-Louis}
\shortauthor{L. Gagliardi \etal}

\institute{                    
  Institut Lumi\`ere Mati\`ere, UMR5306 Universit\'e Lyon 1-CNRS, Universit\'e de Lyon 69622 Villeurbanne, France \\
}
\pacs{91.60.-x}{Physical properties of rocks and minerals}
\pacs{68.55.J-}{Morphology of films}
\pacs{05.90.+m}{Other topics in statistical physics, thermodynamics, and nonlinear dynamical systems}

\abstract{
The forces exerted by growing crystals on the surrounding materials 
play a major role in many geological processes, from diagenetic replacement
to rock weathering and uplifting of rocks and soils.
Although crystallization is a nonequilibrium process, the 
available theoretical prediction for these forces are
based on equilibrium thermodynamics. Here we show that nonequilibrium
effects can lead to a drop of the crystallization force in large pores
where the crystal surface dissociates
from the surrounding walls during growth. The critical pore size
above which such detachment can be observed depends only on the ratio
of kinetic coefficients and cannot be predicted from thermodynamics.
Our conclusions are based on a physical model which accounts for
the nonequilibrium kinetics of mass transport, and disjoining pressure effects 
within the thin liquid film separating the crystal and the surrounding walls.
Our results suggest that the maximum size of the pores that can sustain
crystallization forces close to the equilibrium prediction
ranges from micrometers for salts to  a millimetre for low solubility minerals
such as calcite. These results are discussed in the light of recent experimental observations
of the growth of confined salt crystals.
}

\begin{document} 

\maketitle

\section{Introduction}

The force of crystallization refers to the force that a crystal exerts on the surrounding walls when 
growing in confinement, for instance in a pore of a host material~\cite{Becker1905,Taber1916}. 
These forces play a role in various geological processes. For example 
frost~\cite{Wilen1995,Rempel2001}
or growth of veins in the Earth's crust~\cite{Wiltschko2001,Gratier2012}, 
produce brobdingnagian forces that are able to heave the soil.
They also play
a role in diagenetic replacement~\cite{Maliva1988} and  are one of the major processes involved in 
rock weathering~\cite{Rodriguez-Navarro1999}.
Recently, crystallisation forces have attracted renewed interest due to their role in the weathering 
of buildings and historical heritage~\cite{Schiro2012,Espinosa-Marzal2010,Flatt2014}.

The current understanding of the force of crystallization
relies on equilibrium thermodynamics~\cite{Correns1939,Scherer1999,Steiger2005}.
Some recent experiments have been proposed to  test 
these equilibrium predictions directly and quantitatively~\cite{Desarnaud2013,Desarnaud2016,Naillon2018}. 
Moreover, some theoretical approaches have been proposed to describe nonequilibrium effects~\cite{Naillon2018,Choo2018,Koniorczyk2016}.
However, these approaches do not account for the nonequilibrium processes at play
within the contact, which combine mass transport kinetics and 
physical forces such as 
disjoining pressures and surface tension.
Using a nonequilibrium thin film model to describe the dynamics within contacts, we show that
growing crystals cannot expand their contact regions with the surrounding walls in large pores.
Instead, the crystal surface dissociates from the walls  leading to a drastic drop of the crystallisation force. 
This phenomenon is controlled by a balance between diffusion and precipitation kinetics. 
The critical pore size above which
the force of crystallization vanishes ranges from micrometers for high solubility crystals such as salts, to
the millimetre for low solubility systems such as Calcite.

Our modelling strategy relies on the assumption that a thin liquid film is present in
the contact region between the crystal and the wall. Such a liquid
film can be sustained by disjoining forces when the contact is hydrophilic,
as shown in recent experiments~\cite{Desarnaud2016}.
A liquid film can also be maintained in the contact  
when the substrate is rough, or in the presence of dust particles,
as observed in other experiments~\cite{Kohler2018}.
The evolution of the crystal morphology due to growth or dissolution in the contact 
region is described by the thin film model introduced in Refs.~\cite{Gagliardi2018,Gagliardi2019},
which accounts for disjoining pressure, surface tension, diffusion, and 
surface kinetics.
For the sake of definiteness, we focus on a Hele-Shaw geometry, 
where an axisymmetric crystal grows between two flat and parallel walls
with a purely repulsive disjoining pressure.

We first show that such a model reproduces the expected equilbrium thermodynamic expression 
for the pressure of crystallization~\cite{Correns1939}.
We find that the force is proportional to the area
of contact, and we asseverate the need for a precise 
conventional definition of the contact size
to clarify the discussion on possible correction terms.

In a second part, we focus on non-equilibrium effects.
We model the growth of a crystal with fixed
supersaturation at the edge of the contact region.
We find two different types of dynamics depending
on the value of the dimensionless Darmk\"ohler number, which describes
the competition between surface kinetics and diffusion kinetics. 
For slow surface kinetics, the contact grows and the  nonequilibrium crystallization
pressure is close to the equilibrium prediction. However, for fast enough surface 
kinetics, the crystal surface in the contact detaches from the 
substrate. 
After the detachment, the part of the crystal surface which is still
in contact with the substrate dissolves, and the crystallization force drops and vanishes.

\section{Model}

\begin{figure*}
\center
\caption{
Left panel: Sketch of the axisymmetric model geometry. Section of the crystal along the 
radial coordinate $r$ (solid and dotted blue lines).  
The parallel walls are represented in yellow. 
The dashed grey rectangle delimits the contact region described by the thin film equations. Notations are given in the main text.
Right panel: equilibrium simulations results. 
\textbf{b)} Normalized contact radius estimates, $\bar{L}$, based on three different methods as discussed in the text, and target value $\bar r_{tl}$. 
The size of the simulation box is $\bar{R}_0=40.6$;
\textbf{c)} Normalized equilibrium pressure, $\bar{P}=\bar{F}/(\pi \bar{L}^2)$ versus supersaturation using criterion 
(iii) to compute $\bar{L}$. The dashed line reports the equilibrium expression \cref{eq:force_Correns}.
\label{fig:eq_pressure}}
\includegraphics[width=0.8\linewidth]{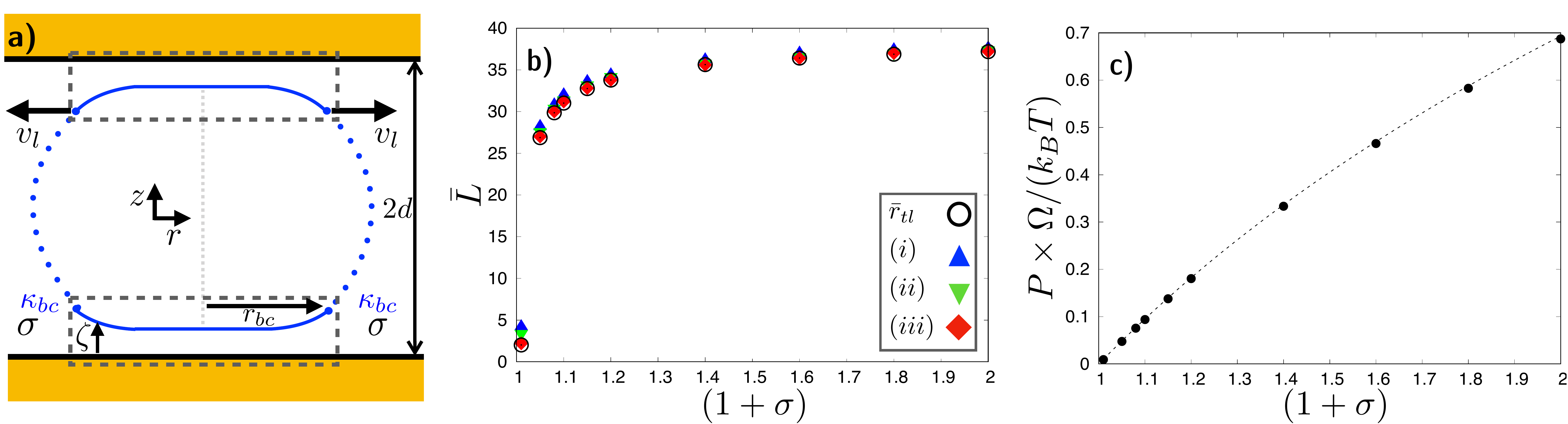}
\end{figure*}

We employ a thin film model~\cite{Gagliardi2018,Gagliardi2018a,Kohler2018,Gagliardi2019} describing the evolution of 
a rigid crystal in a region, hereafter called the contact region, where the crystal is in
the vicinity of a flat substrate.
The model was derived in the small slope limit (also called the lubrication expansion)~\cite{Oron1997}.
Here, we introduce the equations from intuitive physical motivations.
The reader interested in the full derivation of the model equations
should refer to Refs.~\cite{Gagliardi2018,Gagliardi2019}.
In order to simplify the model we consider the 
case of equal densities between the crystal and the liquid. In addition,
we have assumed the dilute limit for the concentration
of crystal ions or molecules in the liquid.

We consider a crystal between two flat walls,
and we focus on an axisymmetric geometry around the $z$ axis, depicted in the left panel of \cref{fig:eq_pressure}. 
Furthermore, we assume the up-down symmetry $z\leftrightarrow-z$, 
so that the two contacts have identical shapes and sizes.
The thickness of the liquid film is denoted as $\zeta(r,t)$,
where $r$ is the radial coordinate, and $t$ is time.
Due to the up-down symmetry, there is no translation of the bulk
of the crystal along $z$, and the local growth rate
of the crystal projected along $z$, $\mathrm v_{z}(r,t)$, is
\begin{equation}
\label{eq:zeta}
\mathrm{v}_z = -\partial_t\zeta\, .
\end{equation}
Moreover, the local concentration
in the contact $c(r,t)$ and $\mathrm v_z$ obey simultaneously two relations 
\begin{subequations}
\label{eq:evolution}
\begin{align}
\label{eq:vz_diffusion}
\frac{\mathrm{v}_{z}}{\Omega}&=\frac{1}{r}\partial_r\left[r\zeta D\partial_r c\right]\, ,
\\
\label{eq:surface_kinetics}
\frac{\mathrm{v}_{z}}{\Omega}&=\nu\left[c - c_{eq}\right]\, ,
\end{align}
\end{subequations}
where $\Omega$ the molecular volume of molecules in the solid,
$D$ is the diffusion coefficient, and $\nu$ the surface kinetic coefficient.
The first relation accounts for diffusion-mediated mass transport along the liquid film.
Remark that the total diffusion mass flux $J_D =-\zeta D\partial_rc$
along the film is proportional to the film thickness $\zeta(r,t)$.
The second equation states that the local growth rate
is proportional to the departure from equilibrium measured by the concentration.
The coefficient of proportionality $\nu$ is called the surface kinetics coefficient. 

The equilibrium concentration $c_{eq}(r,t)$ in \cref{eq:surface_kinetics} describes the concentration
at which attachment and detachment rates compensate, leading to a vanishing growth rate.
When $c=c_{eq}$,
the chemical potential in the liquid $\Delta\mu_L(c)$
is equal to the interface chemical potential $\Delta\mu$,
which accounts for the cost for displacing or deforming the interface
by adding or removing molecules from the solid~\cite{Saito1996,Olivier2016}.
In the dilute ideal limit
\begin{subequations}
\label{eq:chemical_potentials}
\begin{align}
\label{eq:equilibrium_c}
\Delta\mu_L(c) &= k_BT \ln[c/c_0]  \, ,
\\
\label{eq:chem_pot}
{\Delta\mu}&=\Omega\left[\tilde{\gamma}\kappa - U'(\zeta)\right]\, ,
\end{align}
\end{subequations}
where $k_BT$ is the thermal energy, and $c_0$ is a reference concentration (often referred to as the solubility).
The first contribution in $\Delta\mu$  accounts for surface tension effects, and is the product
of the stiffness $\tilde\gamma$ at the orientation of the crystal parallel
to the substrate, with the local curvature $\kappa$.
Within our small slope axisymmetric geometry
\footnote{Two remarks are in order. First, the stiffness tensor is described by a single scalar in the 
small slope axisymmetric geometry. Second, our description based on the
curvature fails in the presence of singular facets. However, such situations can be
handled by a suitable cutoff regularization to achieve a strongly anisotropic behaviour, as discussed in \cite{Kohler2018,Aqua2013,Murty2000}.}, 
we have $\kappa = \partial_{rr}\zeta + \partial_r\zeta/r$.
The second term contains the disjoining pressure $U'(\zeta)$, which is the 
derivative of the interaction potential $U(\zeta)$ between the substrate and the crystal-liquid interface.
In this paper, we will only discuss the case of purely repulsive potentials,
for which $U'(\zeta)<0$. This situation corresponds to 
vanishing macroscopic contact angles.

Using \cref{eq:equilibrium_c,eq:chem_pot},
the relation $\Delta\mu_L(c_{eq})=\Delta\mu$
allows to express $c_{eq}$ as a function of $\zeta$. Inserting this expression into 
\cref{eq:surface_kinetics} and eliminating $c$ between \cref{eq:vz_diffusion,eq:surface_kinetics}
yields an equation for $\mathrm v_z(r,t)$ as a function of $\zeta(r,t)$
\begin{align}
\label{eq:vz}
\frac{\mathrm{v}_{z}}{D}
-\frac{1}{\nu r}\partial_r\left[r\zeta\partial_r \mathrm{v}_{z}\right]
=\frac{\Omega c_0}{r}\partial_r\left[r\zeta\partial_r  
{\rm e}^{[\tilde\gamma\kappa-U'(\zeta)]\Omega/k_BT}\right] \, .
\end{align}
The solution of this equation provides $\mathrm{v}_{z}$,
and the evolution of the local film width $\zeta$ can finally be computed from 
\cref{eq:zeta}.

Once the film thickness $\zeta$ is determined, the force of crystallization 
can be computed as the integral of the disjoining 
pressure over the contact area~\cite{Gagliardi2018}:
\begin{equation}
\label{eq:force}
F =- 2\pi\int_0^{r_{bc}}\!\!\!\! \mathrm{d}r \, r U'(\zeta)\, ,
\end{equation}
where $r=r_{bc}$ is located outside the contact region,
i.e. in a zone where $\zeta(r_{bc})$ is large enough for $U'(\zeta(r_{bc}))$
to be negligible.

\section{Equilibrium}

We first consider the system at equilibrium,
where the chemical potential is equal to a constant denoted as $\Delta\mu_{eq}$.
Using \cref{eq:chem_pot}, we have:
\begin{equation}
\label{eq:mu_eq}
\Delta\mu_{eq} = \Omega\left[\tilde{\gamma}\kappa(\zeta_{eq}) - U'(\zeta_{eq})\right ]\, ,
\end{equation}
where $\zeta_{eq}(r)$ is the equilibrium profile.
Far from the substrate, where the 
disjoining pressure vanishes, the equilibrium profile $\zeta_{eq}(r)$
approaches asymptotically a macroscopic profile $\zeta_{eq}^{\infty}(r)$ 
defined as the solution of \cref{eq:mu_eq} with $U'(\zeta_{eq}^\infty) =0$. This definition implies
that the macroscopic profile  $\zeta_{eq}^{\infty}(r)$ exhibits a constant curvature.
In contrast, in the centre of the contact,
the actual equilibrium profile is roughly flat 
with a vanishing curvature $\kappa(\zeta_0^{eq})\approx 0$ 
and a constant thickness $\zeta_0^{eq} =\zeta_{eq}(r=0)$.
The triple line region is the intermediate
region where the surface profile passes from one of
these asymptotic limiting profiles to the other.
Integrating \cref{eq:mu_eq} and using \cref{eq:force}, we obtain relations
for the macroscopic equilibrium profile $\zeta_{eq}^{\infty}(r)$. A detailed derivation
is reported in Supplementary Material. Evaluating these relations
at an a priori arbitrary position $r=r_{tl}$
inside the triple line region, we obtain two relations.
The first one is 
a radial force balance (along $r$) which accounts for the usual Young-Dupr\'e contact angle relation
in the small slope limit
\begin{equation}
\frac{\tilde{\gamma}}{2}(\partial_r\zeta^\infty_{tl})^2 
= \Delta U +\frac{\Delta\mu_{eq}}{\Omega}( \zeta_{eq}^\infty(r_{tl}) - \zeta^{eq}_0) 
+ \frac{\gamma_{tl}^0}{r_{tl}}\, ,
\label{eq:young}
\end{equation}
with $\Delta U = U(\zeta\rightarrow\infty)- U(\zeta^{eq}_0)$, 
and $\gamma_{tl}^0$ is the triple-line tension neglecting the excess volume
(see Suppl. Mat.).
The triple-line tension is the difference between the free-energy associated to an actual configuration with a straight triple line, 
and that composed of the macroscopic profile with
$\zeta_{eq}^{\infty}(r)$ for $r>r_{tl}$, and $\zeta(r)=\zeta_0^{eq}$ for $r<r_{tl}$.
The usual form of the Young-Dupr\'e relation is retrieved using the small slope relation
$(\partial_r\zeta^\infty_{tl})^2/2\approx \cos(\theta_{eq})-1$, where
$\theta_{eq}$ is the equilibrium contact angle.

The second relation
is a force balance in the direction $z$ orthogonal to the substrate
\begin{align}
F_{eq} &= 
\pi r_{tl}^2\frac{\Delta\mu_{eq}}{\Omega} 
-2\pi r_{tl}\tilde{\gamma} \partial_r\zeta_{eq}^\infty (r_{tl}) \,,
 \label{eq:force_general}
\end{align} 
where the equilibrium force $F_{eq}$ is obtained by inserting the equilibrium profile $\zeta_{eq}(r)$ in \cref{eq:force}.
The terms on the r.h.s. respectively account for the 
cost for changing the thickness of the film within the contact
by adding or subtracting atoms, and for the contribution of
surface tension.

The two relations \cref{eq:young,eq:force} can be used to
describe both the case of partial wetting when $\Delta U>0$ leading to a finite contact angle,
and the non-wetting situation when $\Delta U\leq0$. In the following,
we will focus on purely repulsive potentials with $\Delta U<0$.
In this case, it is convenient to 
choose a definition of the triple-line based on the cancellation
of the macroscopic contact angle
\begin{equation}
\label{eq:der_null}
\partial_r\zeta_{eq}^\infty (r_{tl}) = 0\, .
\end{equation}
Although the physical behaviour is independent of the precise
definition of $r_{tl}$, the expression of the corrections
to the macroscopic limit (terms bringing corrections proportional to the inverse of the size of the crystal
such as the last terms in the r.h.s. of \cref{eq:young,eq:force_general})
will depend on this definition.
Combining \cref{eq:force_general,eq:der_null} the force $F_{eq}$ is found to be proportional
to the contact area $\pi r_{tl}^2$. Then, using \cref{eq:equilibrium_c} 
and the definition of the supersaturation $\sigma=c/c_0-1$,
the equilibrium pressure $P_{eq} = F_{eq}/(\pi r_{tl}^2)$ reads:
\begin{equation}
\label{eq:force_Correns}
P_{eq} = \frac{k_BT}{\Omega}\ln (1+\sigma)\, .
\end{equation}
This expression is identical to that of Correns~\cite{Correns1939}.
However, as opposed to Refs.~\cite{Steiger2005a,Scherer1999},
the pressure of crystallization~\cref{eq:force_Correns}
does not exhibit finite size corrections proportional to the inverse of the size of the crystal.
Such corrections would actually appear if we had chosen a different definition of the
contact radius $r_{tl}$.

\section{Numerical Methods}

\begin{figure}
\center
\caption{Detachment transition.
Blue curve: section of crystal profiles along $r$. We have only represented the two contact regions at the top and at the bottom,
the central part of the crystal between the two plates is not shown.
The vertical scale is enlarged for better visualisation.
\\ a) The crystal shape conforms to the substrate for $\bar{\nu} =10^{-3}$. 
b) Detachment of the crystal surface from the wall for $\bar{\nu} = 4\times 10^{-2}$.
The other parameters are identical: $\sigma = 0.1$, $\bar{d} =100$, and $\bar{\zeta}_{bc} \approx 5.1$. 
\label{fig:profile_animation}}
\includegraphics[width=\linewidth]{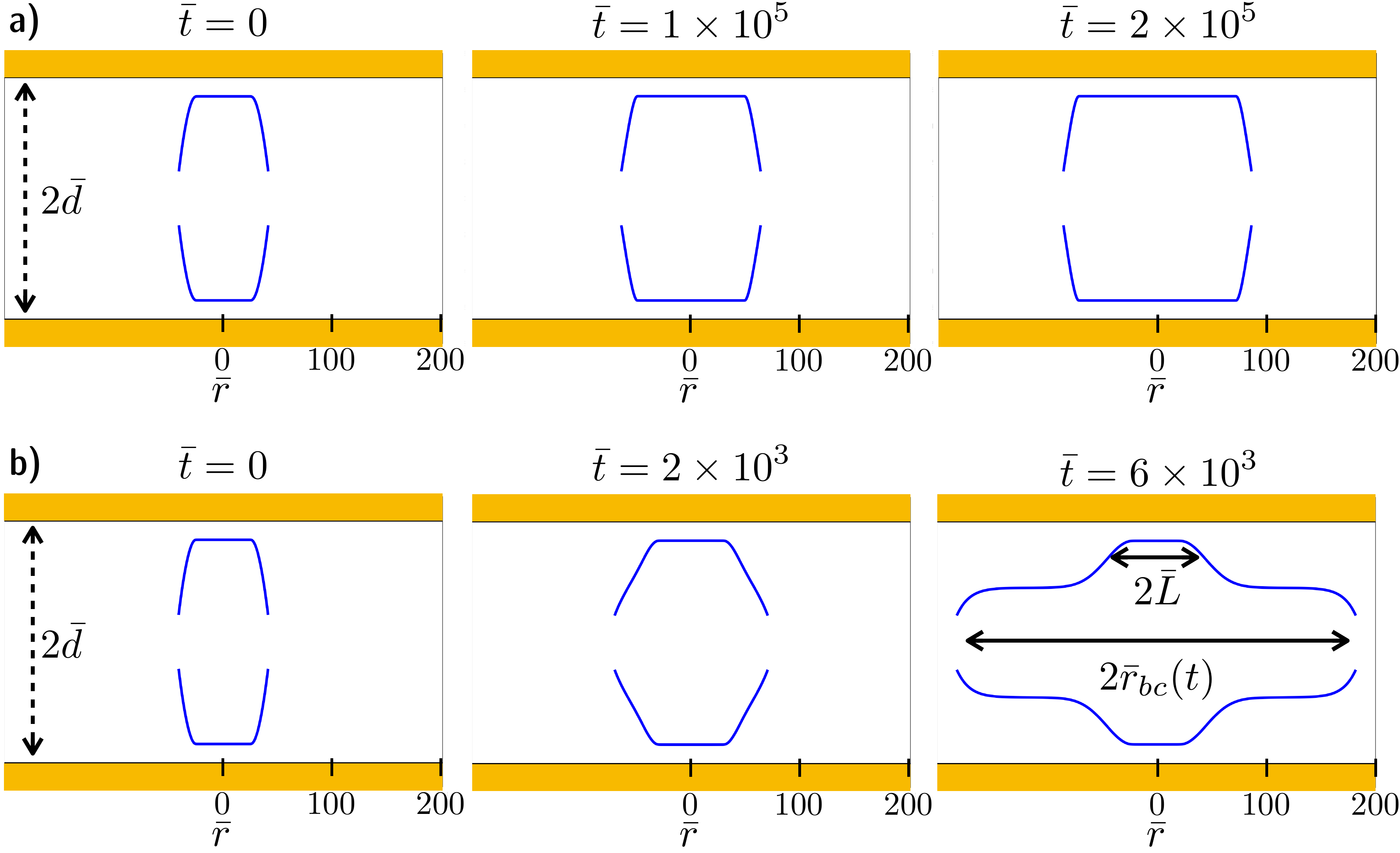}
\end{figure}

We have investigated the equilibrium and non-equilibrium behaviour of our
model using direct numerical simulations of the model equations.
In order to perform simulations, we need to assume a specific form of the 
interaction potential $U(\zeta)$.
We choose a purely repulsive potential, which has proved to provide a 
fair semi-quantitative description of experiments with sodium chlorate crystals~\cite{Kohler2018}
sedimented on a substrate covered with defects or particles of height $h$:
\begin{equation}
\label{eq:potential_CF}
U(\zeta) = A \frac{e^{-(\zeta-h)/(h\bar{\lambda})}}{\zeta-h}\, .
\end{equation}
Here, $\bar{\lambda}h$ is the range of the repulsion.
In our simulations,  the dimensionless repulsion range is fixed to $\bar{\lambda}=0.01$.

The evolution equations \cref{eq:vz,eq:zeta} are solved in a time-dependent 
integration domain of size $r = r_{bc}(t)$. 
At the boundary of the integration domain, we assume a fixed supersaturation 
$\sigma = c(r_{bc})/c_0 -1$ and impose a constant film width $\zeta_{bc}=\zeta(r_{bc})$.
We consider large values of $\zeta_{bc}\gg h$, leading to $U'(\zeta_{bc})\approx 0$.

The surface curvature at the edge of the contact in general depends on the growth dynamics 
outside the contact. Here, we do not solve the dynamics outside the contact. 
Instead, we use the simple assumption of a constant curvature  $\kappa_{bc}=\zeta(r_{bc})$ 
outside the contact.
Such an assumption is consistent with the limit of slow surface kinetics
for an isotropic crystal, where the surface shape is close to the equilibrium 
constant-curvature shape.
 This assumption allows for a straightforward link between the 
curvature and the distance $2d$ between the walls. Two limiting regimes
are considered depending on the value of the radius of the contact $r_{bc}$:
when $r_{bc}\gg d$ the crystal exhibits a disc-like shape
and $\kappa_{bc}\sim 1/d$; when $r_{bc}\ll d$ the crystal shape 
is close to a sphere and $\kappa_{bc}\sim 2/d$.
The results reported below are obtained in the sphere
limit, where $\kappa_{bc}= 2/d$. 
Exploratory simulations for the disc shape indicate that the qualitative behaviour is not affected.

Assuming that the dynamics outside the contact is mainly controlled by
surface kinetics we obtain the velocity $v_l=\dot{r}_{bc}$ at which the edge of the contact expands
\begin{equation}
\label{eq:lateral_vel}
v_l = \left(\partial_r \zeta(r_{bc})\right)^{-1}\left[\Omega\nu c_0\left(1+\sigma-e^{\frac{\Omega\kappa_{bc}}{k_BT}}\right)\right]\, .
\end{equation}

\section{Equilibrium simulations and contact radius}

The numerical determination of the contact radius $r_{tl}$ defined by the relation \cref{eq:der_null}
in general requires the fitting
of the profile outside the contact region, and the evaluation
of the point where the extrapolation of the fitted profile in the 
contact line region exhibits a minimum. 
We wish to design a simpler procedure that would be more convenient,
especially for nonequilibrium simulations.
We therefore examine the accuracy of three possible estimates $L$ of
the contact radius $r_{tl}$:
 (i) $L=\max_r[\zeta''(r)]$; (ii) $L=\min_r[U'''(\zeta)]$; (iii) $L=\max_r[U''(\zeta)]$.

To compare these estimates, we have performed equilibrium simulations.
These simulations were started with a flat contact of size $R_0$. 
For a given supersaturation $\sigma$, we choose $\kappa_{bc}$ such that the equilibrium relation
$\tilde{\gamma}\Omega\kappa_{bc}^0(\sigma)=\Delta\mu_{eq} = k_BT\ln(1+\sigma)$ holds. As a consequence, from \cref{eq:lateral_vel}
$v_l=0$. 
After some short transient dynamical evolution, 
the system reaches equilibrium.
We found no dependence of the results on initial conditions, 
and kinetic parameters ($D$, $\nu$), as expected at equilibrium.

The theoretical value of $r_{tl}$ at equilibrium is obtained from  \cref{eq:force}
as
$r_{tl}=\pi^{-1/2}(F_{eq}\Omega/\Delta\mu_{eq})^{1/2}$. In this expression, the force 
$F_{eq}$ is evaluated
by inserting the equilibrium  profile $\zeta_{eq}(r)$ obtained
from simulations in \cref{eq:force_general}. 
As seen in \cref{fig:eq_pressure}a, different definitions of $L$
disagree only when $L$ is small, i.e. when finite size corrections come into play.
An inspection of \cref{fig:eq_pressure}b reveals that the best estimate is $L = \max_r[U''(\zeta(r))]$.
We will use this definition in the following.
The equilibrium pressure $P_{eq}=F_{eq}/(\pi L^2)$ evaluated with this
definition of $L$ is in good agreement with the Correns expression \cref{eq:force_Correns}, as reported in \cref{fig:eq_pressure}c.
As a consequence, the equilibrium force is fixed by the supersaturation
and does not depend on the expression and parameters of the disjoining potential \cref{eq:potential_CF}.
(Additional results showing that the force is independent of the interaction strength  
are provided in Suppl. Mat.). 

\section{Growth simulations}

\begin{figure}
\center
\caption{
Force, contact area, and pressure of crystallization during growth.
\textbf{a)}: Normalized force of crystallization $\bar F$ as a function of normalized time.
\textbf{b)}: Normalized contact radius $\bar L$, using $L = \max_r[U''(\zeta(r))]$. 
\textbf{c)}: Corresponding normalized non-equilibrium pressures, $\bar{P}=\bar{F}/(\pi\bar{L}^2)$.
In all cases,  $2\bar{d}=200$, $\sigma = 0.1$, $\bar{\zeta}_{bc} \approx 8.1$.
The initial profile was initially equilibrated at $\sigma=0.1$ except for the
violet and green curves in the bottom panel which were equilibrated at $\sigma \approx 0.047$
leading to a different initial profile.
\label{fig:non_eq}}
\includegraphics[width=0.7\linewidth]{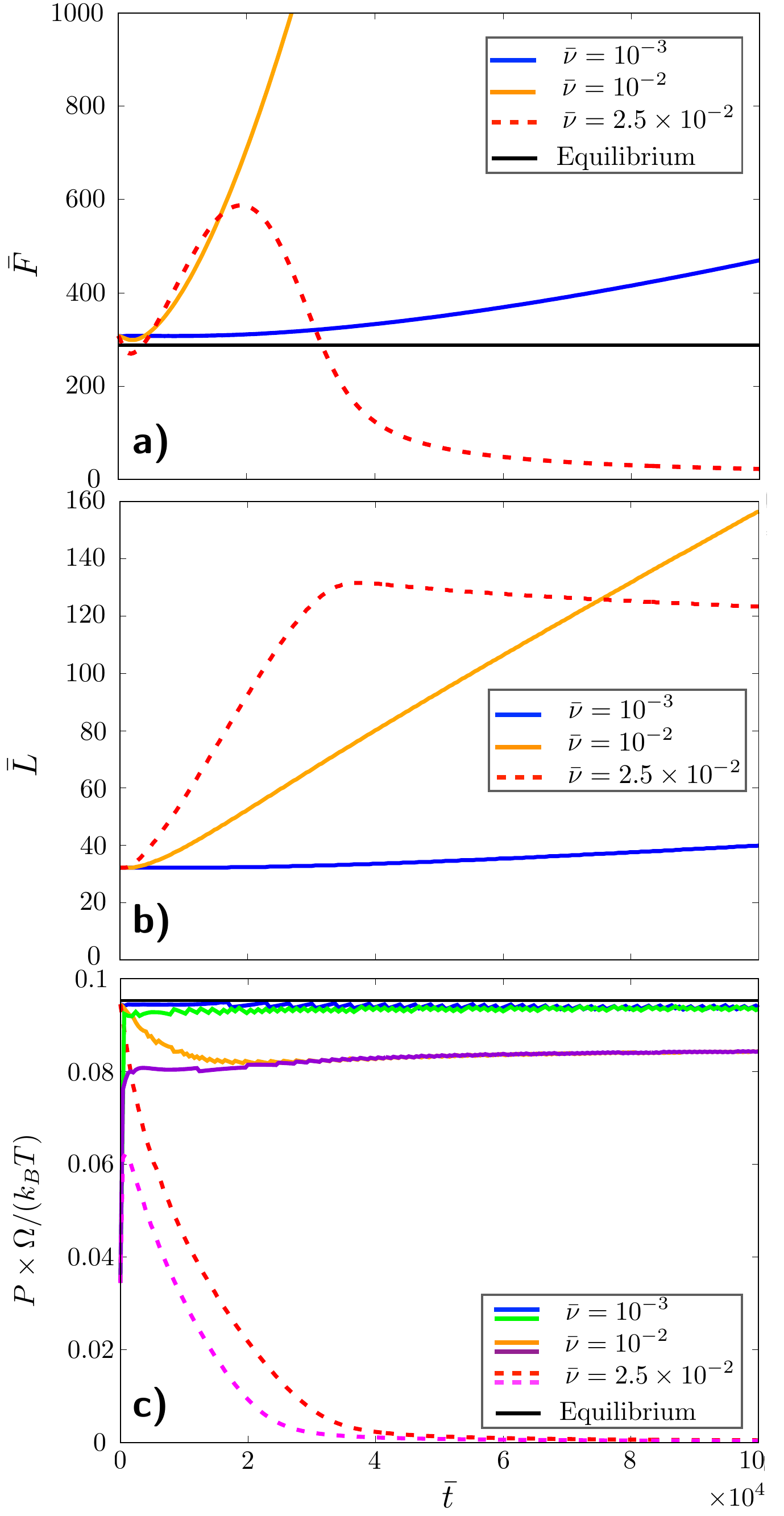}
\end{figure}

Simulations were performed with rescaled dimensionless
model equations. Dimensionless variables are indicated with a bar.
Their detailed definitions are summarised in Suppl. Mat..
Unless specified otherwise, 
the initial profile is an equilibrium profile obtained at $\sigma \approx 0.047$
with a size $\bar{R}_0=R_0/(\Gamma h) = 40.6$,
where $\Gamma=\Omega \tilde\gamma/(k_BT)$. 

Two main regimes are observed depending on $\bar\nu= \nu\Gamma/D$.
For slow surface kinetics, i.e. for small $\bar{\nu}$, 
the crystal profile grows laterally and remains flat in the contact region,
as shown in \cref{fig:profile_animation}a. In this regime,
the width of the film in the contact region is close to the 
equilibrium value $h$. 
In contrast, the crystal detaches from the substrate for faster
interface kinetics, i.e. larger values of $\bar\nu$. This situation is shown in 
\cref{fig:profile_animation}b: after a transient initial growth of the contact region
with film thickness is $\sim h$,
the contact line is pinned, a macroscopic film forms leaving a small contact patch
in the centre. 

These two different regimes give rise to different behaviours
of the crystallization force. For slow attachment kinetics,
the contact radius grows linearly in time, and
the crystallisation force $F$ is proportional to the contact area $\pi L^2$.
As a consequence, the crystallization pressure $P=F/(\pi L^2)$ reaches a constant asymptotically.
This asymptotic value is close to the equilibrium prediction \cref{eq:force_Correns},
as shown in \cref{fig:non_eq}(c)  with $\bar\nu=10^{-3}$
(the small fluctuations of the pressure 
are spurious and are caused by our numerical procedure,
which expands the simulation box on a fixed discretization grid).
For fast attachment kinetics, the crystallization force first increases,
and then decreases to zero after detachment. This behaviour is shown in \cref{fig:non_eq}a
with  $\bar\nu=2.5\times 10^{-2}$. The decrease of the force can be traced back
to the slow dissolution of the remaining contact patch after detachment.
Such a dissolution is due to the excess of chemical potential
of the contact patch, which cumulates high curvature regions and
repulsive disjoining pressures as compared to the rest of the crystal.
The decrease of the size of the contact patch is indeed seen in \cref{fig:non_eq}b.
As the force drops, the equilibrium pressure decreases to zero.
The decrease of the pressure $P=F/\pi L^2$ is a non-trivial statement since
both $F$ and $L$ decrease.
The origin of the decrease of $P$ is the increase of the 
film thickness under the remaining contact patch
during its slow dissolution, which leads to a decrease
of the disjoining pressure $U'(\zeta)$.

Just before the threshold, when $\bar\nu$ is slightly lower than the value 
for which the macroscopic film forms, no detachment transition is observed. 
However, the crystallization pressure  appears to reach a constant which is 
significantly lower than the equilibrium value.
Such a case is reported in \cref{fig:non_eq}
for $\bar\nu=10^{-2}$.
A detailed inspection of the profile $\zeta$ in these simulations
reveals that the film thickness under the crystal indeed reaches a value
which is slightly larger than the equilibrium value,
giving rise to smaller disjoining forces $U'(\zeta)$, and consequently to lower
crystallization pressures.
The long-time behaviour in all regimes is found
to be independent of the initial profile, 
as shown e.g.~by the violet and green curves in \cref{fig:non_eq}c.
Globally, the details of the initial shape are not relevant as long
as a flat contact is present\footnote{Such initial condition 
is not necessarily produced by a growth process and can result
from dissolution, or by squeezing an existing crystal between two plates.}.

\section{Discussion}

\begin{figure}
\center
\caption{Detachment transition.
Normalized film thickness $\bar{\zeta}_f$ at the edge of the contact as a function of the Damk\"{o}hler 
number $\mathcal{D}a=\nu/(D\kappa_{bc})$. 
The vertical dashed line indicates the expected threshold $\mathcal{D}a=1$.
a) $\bar{\zeta}_f$ for various normalized pore sizes $\bar d$, and film widths $\bar{\zeta}_{bc}$ at the edge of the simulated contact region. 
The supersaturation outside the contact zone is fixed to $\sigma = 1$.
b) $\bar{\zeta}_f$ for different values of supersaturation $\sigma$ with fixed $\bar{d}=100$ and $\bar{\zeta}_{bc} =12.5$.
c) (Inset of b) time-evolution of the crystallization pressure above the detachment transition ($\mathcal{D}a = 1.1$) for different supersaturations. 
\label{fig:detachment}}
\includegraphics[width=0.75\linewidth]{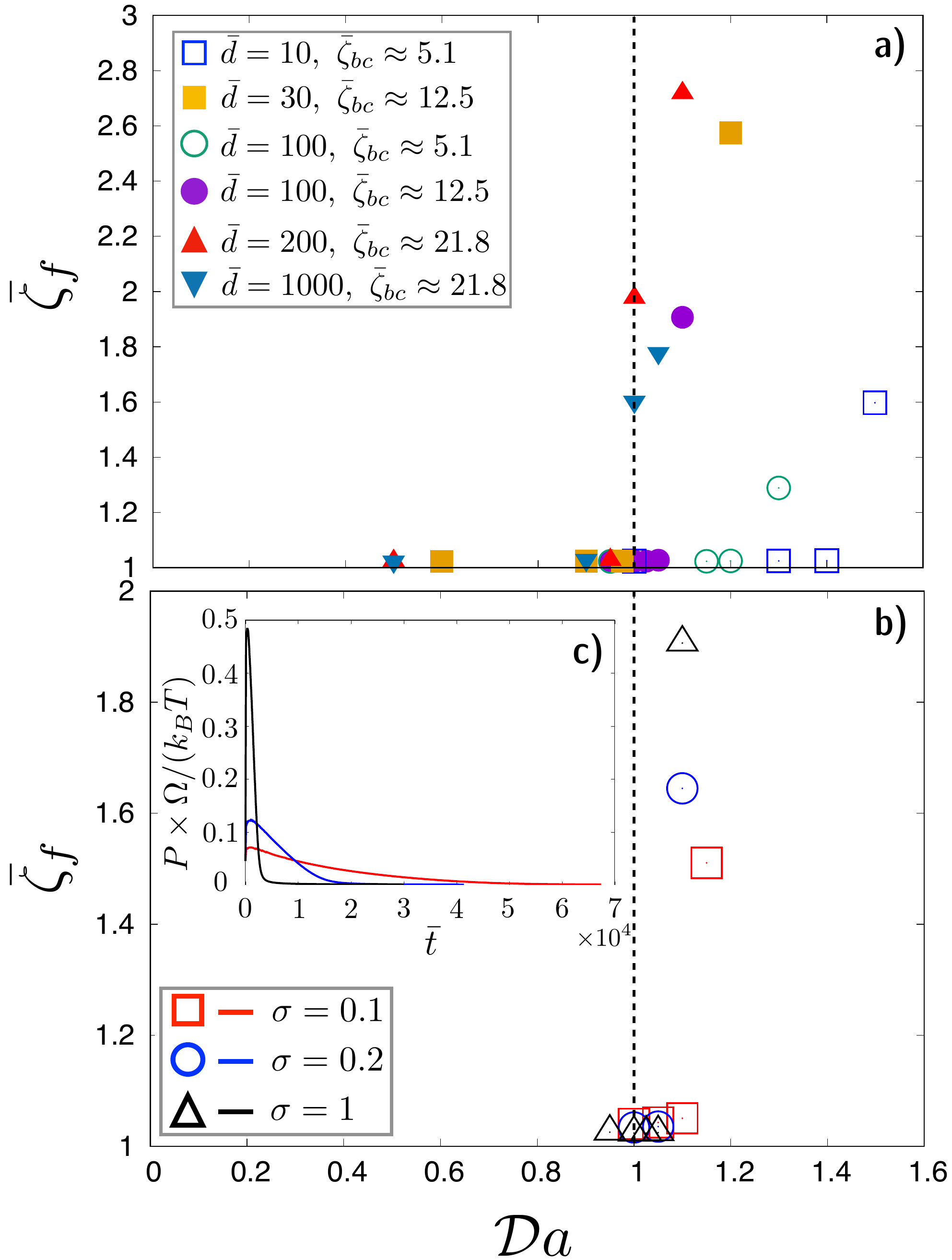}
\end{figure}

The two regimes revealed by simulations can be understood 
from a qualitative discussion of the competition between
diffusion and surface kinetics.
When surface kinetics is slow as compared to diffusion, diffusion 
makes the concentration in the liquid homogeneous.
This leads to a homogeneous supersaturation 
along the surface of the crystal and, in particular, in the contact region. 
Within the contact, this supersaturation  balances the
repulsive part of the disjoining pressure. A local equilibrium
then results in the contact, giving rise to a film thickness and a crystallisation
pressure in agreement with the equilibrium predictions.
In contrast, when surface kinetics is faster,
the growth process occurs so fast that diffusion cannot make
the supersaturation homogeneous. As a consequence, different
parts of the crystal surface are subject to different 
supersaturations. In particular, if a part of the crystal surface
is closer to the source of the supersaturation, which in our case is outside
the contact region, then this part can grow faster than other parts. 
Faster growth outside
the contact region, where the crystal surface is farther
from the substrate, leads to the detachment transition.

The competition between diffusion kinetics and surface kinetics
is quantified by a dimensionless number, usually called the Damk\"{o}hler number~\cite{Fogler2006}. 
For convenience, we use the definition
\begin{equation}
\label{eq:detachment_crit}
\mathcal{D}a = \frac{\nu}{D\kappa_{bc}}\, .
\end{equation}

In order to perform a quantitative analysis of the threshold of the transition,
we measure the film thickness $\zeta_f$,
defined as the thickness at the value of $r$ where  
$\partial_{rr}\zeta(r)$ is minimum.
Before the transition, the curvature is always positive, 
and this corresponds to the film thickness under the crystal where the 
curvature vanishes, ${\zeta}_f \approx h$. After the detachment transition, 
this definition coincides with point where the 
surface of the crystal is the most concave in the zone that connects
the remaining patch and the film (this definition leads to a slight underestimation the thickness 
of the macroscopic film).
The measurement of the film thickness reveals
that the competition between diffusion kinetics and surface kinetics
is characterized by the dimensionless Damk\"{o}hler number introduced above, \cref{eq:detachment_crit}:
indeed, as shown in \cref{fig:detachment}, the transition is found to occur at $\mathcal{D}a \approx 1$.
The results were obtained at large integration times to avoid transient behaviours.
The transition threshold is found to be independent of
the thickness at the boundary $\bar{\zeta}_{bc}$ (\cref{fig:detachment}a). 
Additional numerical results shown in \cref{fig:detachment}b
confirm the negligible effect of the supersaturation at the boundary of the contact region on the transition.
However, the supersaturation affects significantly the relaxation time to reach the asymptotic pressure. 
In particular, as illustrated by \cref{fig:detachment}c, the decrease of $P$ above the transition is slower as  $\sigma$ is decreased.
We wish to stress on the fact that the criterion for the detachment transition $\mathcal{D}a>1$, 
which leads to the drop of the crystallisation force, is 
not only independent of supersaturation, but also 
independent of the details of the 
repulsion potential (as long as the disjoining pressure is repulsive), and of surface tension.
This is purely a kinetic balance.

In the Hele-Shaw geometry discussed here,  $\kappa_{bc}\sim 1/d$ 
and the detachment transition should appear for $\mathcal{D}a=d\nu/D\sim 1$.
In the case of salt, 
the reported kinetic constants span a large range of values~\cite{Colombani2012},
from $10^{-5}$ to $10^{-3}$ms$^{-1}$. 
Assuming $\nu= 10^{-3}\mathrm{m\,s^{-1}}$~\cite{Naillon2017,Naillon2018}
and $D\approx 10^{-9}\mathrm{m^2s^{-1}}$~\cite{Chang1985}, 
we obtain that the critical pore size above which a drop in the (nonequilibrium) 
crystallization force should be observed, is $2d\approx 1\mathrm{\mu m}$.
Such an order of magnitude can be discussed within the frame of
recent crystallization force experiment by Naillon et al.~\cite{Naillon2018}.
These authors found an extremely small crystallization pressure 
in channels with square cross-section $5\times5\mu\mathrm{m}$ (and channel lengths $>100\mathrm{\mu m}$).
However, their interpretation of this drop is based on the decrease of the supersaturation
in the vicinity of the crystal due to the limitation of the diffusion-mediated transport
in the channel far from the crystal. In contrast, our claim is that even 
in the absence of any drop of the supersaturation in the vicinity of the 
crystal, the crystallization force should drop.
Furthermore, to a first approximation, the effect of this drop could be accounted for by a slow decrease of the supersaturation $\sigma$ in the vicinity of the crystal. Since, as discussed above, the instability threshold is independent of $\sigma$, we expect that this slow decrease will not affect the instability threshold.
However, a precise understanding of the effect of the boundary
conditions for a given geometry would require to solve the diffusion field not only
in the contact region, but also outside it.

Experiments of halite growth in a $600\times100\mu$m
PMDS channel reported by Sekine et al~\cite{Sekine2011} found
an inhomogeneous force localized at rims emerging from facet corners.
The differences between these observations and our model 
could originate in the softness or permeability of the PDMS substrate.

In other recent experiments, Desarnaud et al.~\cite{Desarnaud2016} grew salt crystals between two glass plates 
with a larger separation $2d\approx 50\mathrm{\mu m}$.
These experiments produce a pressure of crystallization consistent with the equilibrium formula. 
Since in these experiments the crystals reached zero lateral growth, equilibrium might 
have been reached before the detachment transition.
These results again call for a detailed analysis of the time-dependence of
the supersaturation at the boundary of the contact region.

Finally, experiments have also been performed with calcite crystals~\cite{Li2017} which are much less soluble than salts. 
Using $\nu=10^{-6}\mathrm{m\,s^{-1}}$~\cite{Colombani2016} and $D\approx 10^{-9}\mathrm{m^2s^{-1}}$~\cite{Li1974}, 
the critical pore size rises to 
$2d\approx 1$mm. Hence, calcite crystals could
produce large crystallization forces even in large pores.
However, a quantitative discussion is difficult due to the 
uncertainty in the quantitative experimental measurements of $\nu$~\cite{Colombani2016}.

The detachment transition itself is a consequence of diffusion-limited growth,
which produces other well-studied instabilities such as the Mullins-Sekerka instability giving rise to dendrites, 
or Hopper growth. Hopper growth with salt crystals, giving rise to very small 
contact areas between the crystal and the surrounding walls has been recently 
observed in capillaries of width $\sim 100\mu$m~\cite{Desarnaud2018}.
These results suggest that 
complex morphological instabilities should come into play
when the width of the pores is increased beyond the value for which the detachment transition occurs.

\section{Conclusions} 

In this work we have investigated the nonequilibrium force of crystallization exerted by a 
growing crystal between two parallel walls.
Assuming that a liquid film is maintained by repulsive disjoining forces between the crystal and the walls, 
we showed that two main nonequilibrium regimes are expected. 
When surface kinetics is slow, the crystal surface is homogeneously supersaturated,
and the growth conforms to the shape of the substrate.
The crystallization pressure is then close to the equilibrium value fixed by the imposed supersaturation.
In contrast, when surface kinetics is faster as compared to transport by diffusion,
the supersaturation becomes inhomogeneous. This favours faster growth outside the contact,
and leads to a detachment transition accompanied with a drop of the crystallisation force.
Our results suggest that crystals with fast surface kinetics cannot sustain large forces of crystallization
when growing in large pores.

\acknowledgments
This project has received funding from the European Union's Horizon 2020 research and innovation programme under the Marie Sklodowska-Curie Grant Agreement No. 642976 (ITN NanoHeal).

\bibliographystyle{eplbib.bst}
\bibliography{biblio_CF}

\end{document}